\documentclass[twocolumn,aps,superscriptaddress,10pt,dvips,showpacs,showkeys,prl]{revtex4}
\usepackage[utf8]{inputenc}
\usepackage{graphicx}
\usepackage{setspace}
\usepackage{units}
\usepackage{xspace}
\usepackage{color}

\begin{document}

\title{Mn-induced modifications of Ga $3d$ photoemission from (Ga,Mn)As: evidence for long range effects} 
\author{J. Kanski} 
\email{janusz.kanski@chalmers.se}
\affiliation{Department of Experimental Physics, Chalmers University
  of Technology, SE-412 96 G\"oteborg, Sweden}
\author{I. Ulfat} 
\email{intikhab.ulfat@chalmers.se}
\affiliation{Department of Experimental Physics, Chalmers University
  of Technology, SE-412 96 G\"oteborg, Sweden}
\author{L. Ilver} 
\email{lars.ilver@chalmers.se}
\affiliation{Department of Experimental Physics, Chalmers University
  of Technology, SE-412 96 G\"oteborg, Sweden}
\author{M. Leandersson} 
\email{janusz.sadowski@maxlab.lu.se>}
\affiliation{MAX-lab, Lund University,
 SE-221 00 Lund, Sweden}
\author{J. Sadowski} 
\email{janusz.sadowski@maxlab.lu.se>}
\affiliation{MAX-lab, Lund University,
 SE-221 00 Lund, Sweden}
\author{K. Karlsson} 
\email{krister.karlsson@his.se>}
\affiliation{Department of Life Sciences, University of Sk\"ovde,
 SE-541 28 Sk\"ovde, Sweden}
\author{P. Pal} 
\affiliation{National Physical Laboratory, New Delhi 110012, India}

\date{\today}
\begin{abstract}
Using synchrotron based photoemission, we have investigated the Mn-induced changes in Ga $3d$ core level spectra from as-grown ${\rm Ga}_{1-x}{\rm Mn}_{x}{\rm As}$. Although Mn is located in Ga substitutional sites, and does therefore not have any Ga nearest neighbours, the impact of Mn on the Ga core level spectra is pronounced even at Mn concentrations in the range of 0.5 \%. 
The analysis shows that each Mn atom affects a volume corresponding to a sphere with around 1.4 nm diameter. 
\end{abstract}
\maketitle

(Ga,Mn)As is by far the most intensely studied dilute magnetic semiconductor. Despite the continued interest since its first synthesis \cite{Ohno} and despite successful demonstrations of proof-of-concept devices \cite{Chun,Tanaka}, practical implementations remain unrealistic because the Curie temperature is still below 200 K \cite{Olej,Chen}. One hurdle in the development has been a lack of understanding of the nature of the ferromagnetic state. While it is clear that the ferromagnetism is carrier mediated \cite{Dietl}, it is not known which electronic states are actually the ones mediating the coupling. Different models are discussed, based on different coupling mechanisms. The so far most successful description, the RKKY model is based on spin polarization of holes in the valence band of the host material. More recently, studies of transport, thermoelectric, and magnetic dependence of hole concentration \cite{Ma} have provided support for the impurity band model, in which the alignment of magnetic moments is obtained through double exchange mediated by hopping between impurity states. A third, less discussed model, ascribes the development of the ferromagnetic state to percolation of magnetically ordered regions, described as bound magnetic polarons \cite{Kaminski}.  
A key issue in this context is to establish the range of interaction of an individual Mn atom in the GaAs lattice. In the present study we address this question by means of core level photoemission, a well-established method for probing the changes in the local potential at atomic sites surrounding a chemical impurity. The ferromagnetically active Mn atoms occupy Ga sites in the GaAs lattice, so it is natural to expect the largest impact of Mn on the nearest neighbor As atoms. However, as the samples are prepared by low temperature MBE, the surfaces are inevitably terminated with As in different states, making the analysis of As core level spectra from (Ga,Mn)As quite complicated \cite{Uf1}. Therefore we focus here on the analysis of the Ga $3d$ spectra, which are accessible at the photon energies available at BL I3. From previous experiments we know that the Ga $3d$ emission is strongly modified in (Ga,Mn)As in comparison with GaAs, but no special effort has been made earlier to characterize these effects.

The experiments were carried out at BL I3 at the Swedish National Synchrotron Radiation Laboratory MAX-lab. The beamline is using a normal incidence monochromator, primarily designed for high resolution studies at energies up to around 50 eV photon energy and the photoemission spectra are recorded with a Scienta R4000 hemispherical analyzer. The samples were produced in a SVTA MBE system, directly connected to the spectrometer to allow transfer of as-grown samples under UHV conditions. Samples with different Mn concentration were grown on epi-ready $n$-doped GaAs (100). After thermal desorption of the native oxide, a GaAs buffer was deposited at a growth rate of 0.2 ML$s^{-1}$, with the substrate at 580 $^{\circ}$C. The substrate temperature was then reduced to a value in the range 180-280 $^{\circ}$C depending on the intended Mn concentration, whereupon an LT GaAs buffer and a (Ga,Mn)As layer was grown. A motor controlled substrate shutter was used to leave a part of the LT GaAs surface free from (Ga,Mn)As, so that the clean GaAs could serve as a reference in the following studies. Throughout the preparation the growth rate and surface quality was monitored by means of RHEED. 

\begin{figure}[t]
\includegraphics[height=5cm]{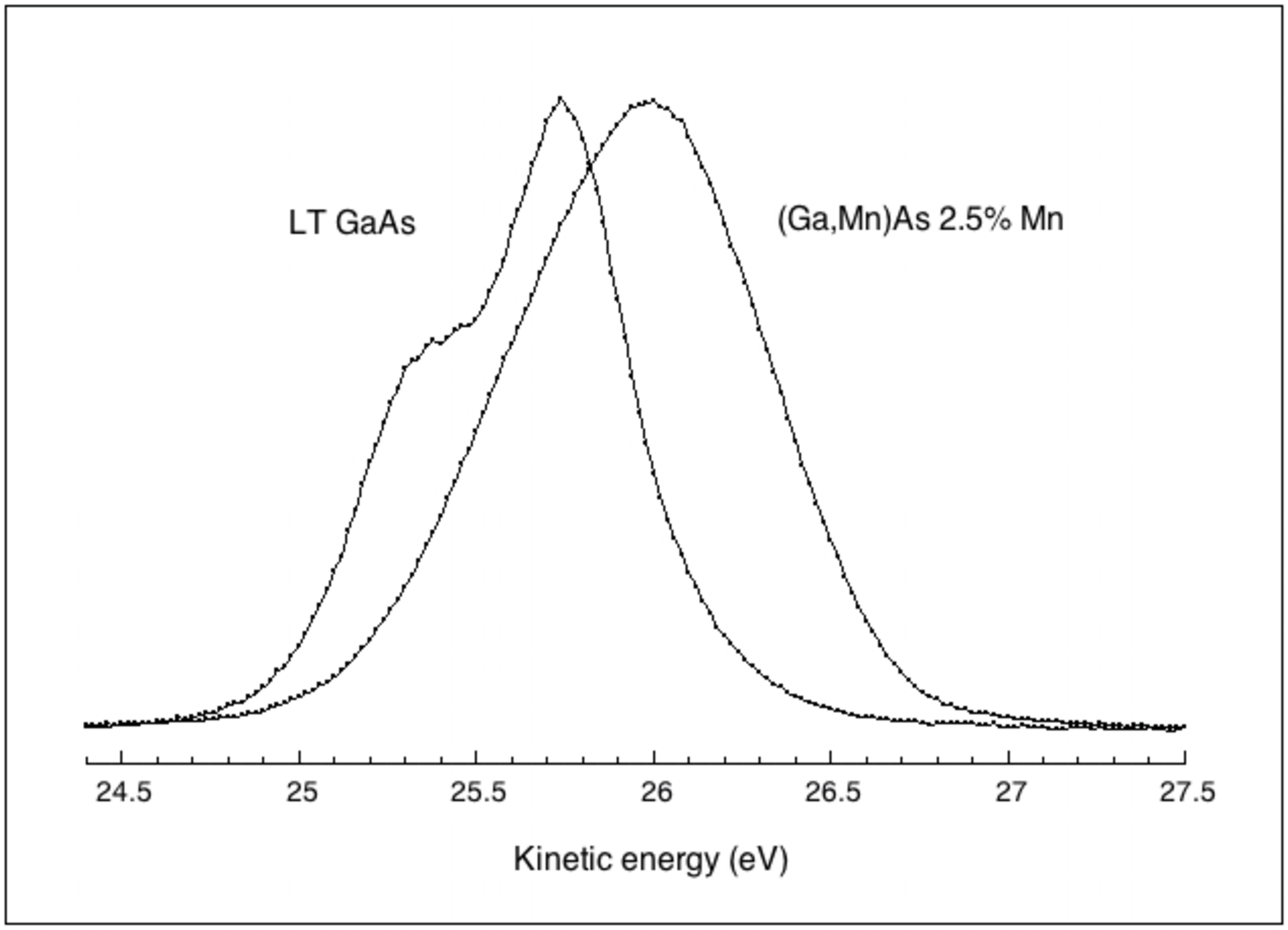}
\caption{
Ga $3d$ spectra from a (Ga,Mn)As(100)-(1x2) sample with 2.5 \% Mn and a reference GaAs(100)-c(4x4) sample. The spectra were excited with 49.5 eV photon energy. The relative shift in energy between the two spectra is mainly due to different Fermi level pinnings. 
}
\label{Ga3d-fig1}
\end{figure}

\begin{figure}[t]
\includegraphics[width=\columnwidth, bb=0 50 595 842]{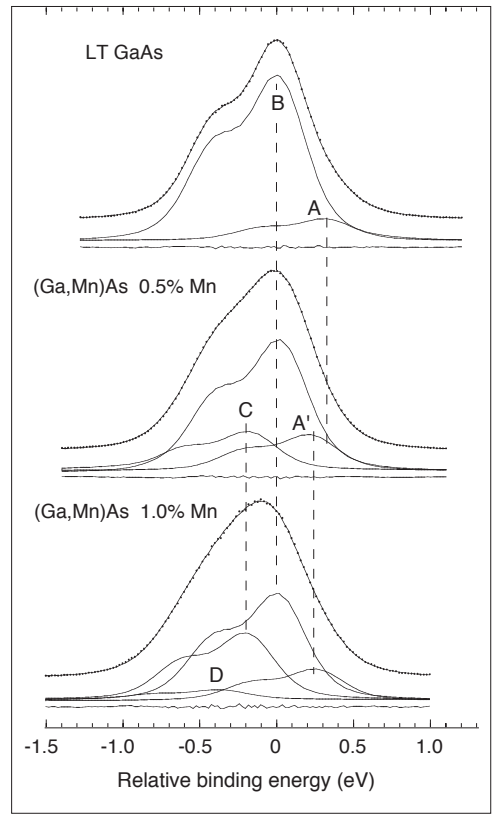}
\caption{
Ga $3d$ spectra from clean GaAs(100) (top), (Ga,Mn)As with 0.5 \% Mn (middle), and with 1 \% Mn (bottom). The spectra are shifted in energy to align the main components, which are assumed to represent Ga atoms unaffected by Mn. 
}
\label{Ga3d-fig2}
\hfill
\end{figure}

As already mentioned, the Ga $3d$ spectra are known to be sensitive to the introduction of a few \% Mn into GaAs, despite the fact that the Ga and substitutional Mn atoms are not nearest neighbors in the lattice. The situation is exemplified in Fig. \ref{Ga3d-fig1}, where we show the spectra from LT GaAs and (Ga,Mn)As with 2.5\% Mn. Two types of changes are observed: a shift and a broadening. The shift is at least partly explained by the different Fermi level positions in the two cases: in LT GaAs the Fermi level is located near midgap, while in (Ga,Mn)As it is at the VBM. The broadening can be a sign of an inhomogeneous potential distribution within the probed area. However, as the (Ga,Mn)As is strongly $p$-doped, the electrostatic potential should be quite uniform. This is in fact confirmed by valence band spectra, where structures reflecting critical points (e.g. the $X_3$ point) do not show any sign of broadening \cite{Uf2}. Therefore we assume that the broadening is an effect of new components appearing in the Ga spectra of (Ga,Mn)As. 
Detailed analysis of a spectrum like that from the (Ga,Mn)As sample in Fig. \ref{Ga3d-fig1} is quite arbitrary without more detailed knowledge about the system, e.g. the expected number of inequivalent atomic sites. In the present case we can assume that the additional spectral component(s) are not surface related, since the surface is terminated with As. This assumption can be verified by comparing spectra that are recorded under conditions with different surface sensitivities, as demonstrated below. Apart from this, the number of inequivalent sites is not known a priori. A way to handle this complication is by following the development of the spectrum from very low Mn concentrations. For this reason we have studied samples with Mn concentrations in the range 0.1 - 1\%.  Fig. \ref{Ga3d-fig2} shows the Ga $3d$ spectra from three samples, with 0, 0.5, and 1.0 \% Mn. All spectra shown were excited with 49.5 eV photons and recorded in normal emission. The figure also shows the results of spectral decomposition obtained with a curve fitting routine (Rainbow) using the following fitting parameters: spin-orbit splitting $\Delta E_{SO}$ = 0.42 eV, branching ratio is 1.82, Gaussian broadening $\Delta E_G$ = 0.318 eV, Lorentzian broadening  $\Delta E_L$ = 0.181 eV, and an exponential background. The quality of each fitting can be estimated from the difference curve between the experimental and the fitted data, shown below each set of curves. For comparison the spectra are aligned in energy with respect to the main peak (B), which is assumed to represent Ga atoms not affected by Mn impurities. Positive values on the scale represent higher kinetic energy than the main component.  
The spectrum from LT GaAs contains two components. As discussed earlier\cite{Asklund}, the high-energy component (A) can be associated with Ga atoms coordinated to As atoms around an As antisite defect. Its relative magnitude depends on growth temperature and does not depend on the probing depth. In the present case its relative area is around 12 \%, in good agreement with corresponding data (250 $^{\circ}$C growth temperature) in \cite{Asklund}. Considering that each As antisite defect has 12 Ga atoms around it, we can conclude that around 1\% of all Ga sites are occupied by an As atom in this case. 

\noindent

\begin{figure}[t]
\includegraphics[width=\columnwidth, bb=0 150 595 842]{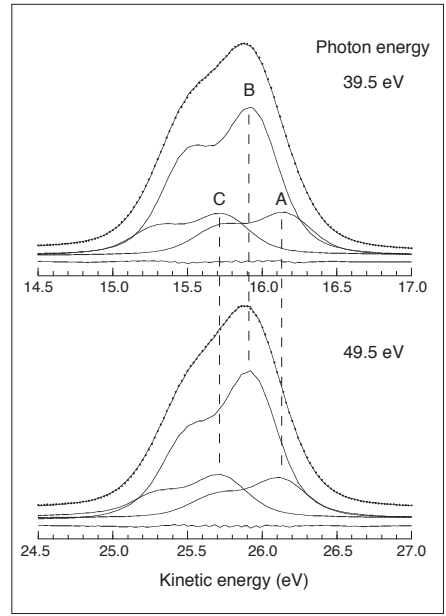}
\caption{
Ga $3d$ spectra from the sample with 0.5 \% Mn, excited with 39.5 eV (top) and 49.5 eV photons (bottom).
}
\label{Ga3d-fig3}
\end{figure}

Turning to the 0.5 \% spectrum (middle spectrum in Fig. \ref{Ga3d-fig2}), we see that it is already clearly affected by the presence of Mn. One can still observe the spin-orbit related asymmetry, which facilitates the spectral decomposition. A high-energy component (A$'$) is found, just as for clean GaAs, but its relative magnitude is increased to around 18 \%, and its energy separation relative to the main peak is reduced by 0.14 eV. Since a very good fit was obtained with the same set of parameters as for clean GaAs without invoking any new high-energy component, it is likely that the increased intensity of (A$'$) relative (A) is due to an increased density of As antisites. Noting that the two samples were parts of the same substrate and grown under identical conditions, we can exclude that the changed intensity is due to reduced growth temperature (as in \cite{Asklund}). There is another mechanism by which the density of As antisites might be increased, namely as an effect of self compensation (i.e. compensation of Mn acceptors by generation of As antisite donors). Such effects have been reported in connection with modulation-doped hetero-structures \cite{Yu}, in which case the density of Mn interstitials (which act as double donors just like As antisites) was increased with increased $p$-type doping. The shift relative the main component can be ascribed to the fact that (Ga,Mn)As is highly $p$-doped, and therefore both the initial and final state energies of the two types of sites can be affected in slightly different ways. A detailed analysis of the shift is beyond the scope of this work. We turn now to the focus of the present study, the low energy region of the spectrum where a completely new component (C) is found. This component is located at -0.20 eV relative the main peak and its magnitude is about 17 \% of the total intensity. The bulk origin of this component (as well as that of the high energy component) is demonstrated in Fig. \ref{Ga3d-fig3}, where we show spectra excited with 49.5 eV and 39.5 eV photon energies. The corresponding mean free paths in the two cases are around 7 and 15 {\AA} \cite{Zangwill}, respectively, so the different surface sensitivities would give significantly modified intensity ratios between the two spectra for a surface related component. The similarity between the two spectra assures that all features are bulk derived. 
Due to the rapidly varying excitation cross section of the Ga $3d$ states, different branching ratios had to be used in the fittings at the two photon energies. The other fitting parameters, including the relative positions of the components were, however, identical. 

Since we know that the density of Mn atoms is around 0.5 \%, and the fraction of affected Ga atoms is larger than that connected with As antisites, we can conclude that the range of interaction of a Mn impurity is larger than that of an As antisite. Due to the relatively large uncertainty of the Mn concentration (very similar spectra were obtained with nominal Mn concentrations down to 0.1 \%), we proceed to the last spectrum in Fig. \ref{Ga3d-fig2}, representing a nominal Mn concentration of 1 \%. 
The spectrum from the 1 \% sample is even more broadened, still asymmetric, but the shoulder reflecting the $3d_{3/2}$ emission is hardly visible. The high-energy component (A$'$) appears at nearly the same energy as for 0.5 \% (at +0.24 eV). Its relative intensity is about 15 \%, similar to that of the more dilute sample, which supports the above interpretation of the increased intensity relative the clean GaAs case. On the low energy side the additional peak (C) appears at precisely the same energy (-0.20 eV) as in the previous case. However, its intensity is markedly increased, reaching about 31 \% of all Ga atoms (including those coordinated to As antisites). To obtain a good fit with the same set of parameters, a fourth spectral component (D) had to be included, located at -0.39 eV. The relative intensity of this component is slightly below 5 \%. 
As the concentration of Mn is only around 1 \% (with an accuracy better than 0.5 \%), the large fraction of Ga atoms represented by component (C) shows immediately that this component cannot be explained by direct bonding interaction. To quantify the volume affected by each Mn impurity, we can relate to the unit cell of the zincblende  structure of (Ga,Mn)As. Since each unit cell contains 4 atoms, 1 \% Mn concentration means that there is 1 Mn atom per 25 unit cells. About 30 \% of this volume is affected by the single impurity, i.e. around 8 unit cells. In other words, an individual Mn impurity affects all Ga atoms within a sphere of 1.4 $\pm$ 0.2 nm diameter (using the mentioned uncertainty in Mn concentration), and all atoms within this volume are affected in the same way. Regarding the small component at -0.39 eV, we note that according to a Monte Carlo simulation \cite{Uf3} the probability of finding two Mn atoms within the distance of interaction is around 6 \% for a Mn concentration of 1 \%. This corresponds quite well to the relative intensity of component (D). It is easy to see that further broadening associated with new spectral components can be anticipated at higher Mn concentrations. As the number of combinations is rapidly increasing, analysis of spectra from samples with higher Mn concentrations becomes a very intricate problem. 

Finally we address the question concerning the origin of the new component in the Ga $3d$ spectrum. Core level shifts in photoemission are generally caused by several simultaneously acting mechanisms that can be associated with initial- and final state phenomena. Without computational analysis it is usually not possible to assess the relative importance of the different effects. Yet, a qualitative interpretation based on general arguments can often be made. In the present case we find that a single Mn atom affects a large number of Ga sites, and the effect is independent of the distance from the Mn atom. These observations exclude local mechanisms like chemical interaction as a source of the observed shift. 
An obvious difference between pure GaAs and (Ga,Mn)As is the presence of free carriers in the latter due to Mn-induced $p$-doping. One consequence of the doping is a shift of the Fermi level, which results in increased kinetic energies of the photoemitted electrons. This alone obviously cannot explain the evolution of a second component in the Ga spectrum. However, by considering the detailed nature of Mn acceptors, we find a possible mechanism that can explain the present observations. It is quite established that Mn atoms in Ga substitutional sites are divalent, occurring as either neutral ($3d^{5}$ + hole) or ionized $3d^{5}$ acceptors. The ionization energy is 113 meV \cite{Schneider}, significantly larger than the value predicted by effective mass calculations \cite{Balder}. This means that at room temperature the majority of the Mn acceptors is in the neutral state (although we note that electron paramagnetic resonance (EPR) data suggest that only the ionized acceptors are present even at 3.4 K \cite{Tward}). Assuming that the bound hole state prevails dynamically (on the time scale of photoemission), Ga atoms within its range can be expected to be distinguishable in core level spectra as a result of both initial state (positive potential) and final state (screening) effects. With an effective heavy hole mass at the valence band maximum of 0.45 $m_0$ and a binding energy of 113 meV, the effective Bohr radius of the bound hole would be 6 \AA, strikingly similar to the range of interaction obtained from the present core level data. Although the close similarity may be incidental, it is nevertheless compelling. We thus propose that the Mn-induced component in the Ga $3d$ spectra found here reflects emission from Ga atoms within the range of the bound hole.

In summary, the detailed analysis of Ga $3d$ spectra from dilute (Ga,Mn)As shows that  each Mn impurity affects the electronic properties of its surrounding in a volume within a radius of 0.7 $\pm$ 0.1 nm, most likely via a bound carrier state. This resembles the scenario described by the bound magnetic polaron model \cite{Kaminski}. Specifically, the decay length of the hole wave function in (Ga,Mn)As matches well the range of interaction deduced in the present study.



\begin{thebibliography}{99}                                                                                               %
\bibitem{Ohno} H. Ohno, A. Shen, F. Matsukura, A. Oiwa, A. Endo, S. Katsu-moto, and Y. Iye, Appl. Phys. Lett. \textbf{69}, 363 (1996). 
\bibitem{Chun} S. H. Chun, S. J. Potashnik, K. C. Ku, P. Schiffer, and N. Samarth, Phys. Rev. B \textbf{66}, 100408 (2002).
\bibitem{Tanaka}	M. Tanaka and Y. Higo, Phys. Rev. Lett. \textbf{87}, 026602 (2001).
\bibitem{Olej} K. Olejnik, M. H. S. Owen, V. Novak, J. Masek, A. C. Irvine, J. Wunderlich, and T. Jungwirth, Phys. Rev. B \textbf{78}, 5, 054403 (2008).
\bibitem{Chen} L. Chen, S. Yan, P. F. Xu, J. Lu, W. Z. Wang, J. J. Deng, X. Qian, Y. Ji, and J. H. Zhao, Appl. Phys. Lett.  \textbf{95}, 182505 (2009).
\bibitem{Dietl} T. Dietl, H. Ohno, and F. Matsukura , Phys. Rev. B \textbf{63}, 195205 (2001). 
\bibitem{Ma} P. Mahadevan and A. Zunger, Appl. Phys. Lett. \textbf{85}, 2860 (2004).
\bibitem{Kaminski} A. Kaminski and S. Das Sarma, Phys. Rev. Lett. \textbf{88}, 247202 (2002).
\bibitem{Uf1}	I. Ulfat, J. Adell, J. Sadowski, L. Ilver, and J. Kanski, Surf. Sci. \textbf{604}, 125 (2010).
\bibitem{Uf2}	I. Ulfat, J. Kanski, L. Ilver, K. Karlsson, and M. Leandersson, to be published.
\bibitem{Asklund} H. {\AA}sklund, L. Ilver, J. Kanski, J. Sadowski, and M. Karlsteen, Phys. Rev. B \textbf{65}, 115335 (2002).
\bibitem{Yu}	K. M. Yu, W. Walukiewicz, T. Wojtowicz, W. L. Lim, X. Liu, M. Dobrowolska, and J. K. Furdyna, Appl. Phys. Lett. \textbf{84}, 4325 (2004).
\bibitem{Zangwill}	A. Zangwill, Physics at Surfaces, Cambridge University Press, Cambridge (1998).
\bibitem{Uf3} I. Ulfat, J. Kanski, L. Ilver, J. Sadowski, K. Karlsson, A. Ernst, and L. Sandratskii, to be published.
\bibitem{Schneider} J. Schneider, U. Kaufmann, W. Wilkening, and M. Baeumler, Phys. Rev. Lett. \textbf{59}, 240 (1987). 
\bibitem{Balder} A. Baldereschi and N.O. Lipari, Phys. Rev. B \textbf{9},1525 (1974).
\bibitem{Tward} A. Twardowski, Materials Science and Engineering B \textbf{63}, 96 (1999).
\end{thebibliography}
\bibliographystyle{apsrev}
\end{document}